\documentclass[aps,prb,reprint,superscriptaddress]{revtex4-2}

\usepackage{graphicx}
\usepackage[colorlinks=True,linkcolor=red,citecolor=blue,urlcolor=blue]{hyperref}
\usepackage[version=4]{mhchem}
\usepackage{siunitx}
\usepackage{multirow}
\usepackage{bm}
\usepackage{amssymb}
\usepackage{array}
\usepackage{makecell}
\usepackage{braket}
\usepackage{booktabs}
\usepackage{dcolumn}

\newcommand{\unit}[1]{\,\mathrm{#1}}
\begin{document}

\title{Strong anisotropy of electron-phonon interaction in NbP probed by magnetoacoustic quantum oscillations}

\author{Clemens Schindler}
\email{clemens.schindler@cpfs.mpg.de}
\affiliation{Max Planck Institute for Chemical Physics of Solids, 01187 Dresden, Germany}
\affiliation{Institut f\"{u}r Festk\"orper- und Materialphysik, Technische Universit\"at Dresden, 01062 Dresden, Germany}

\author{Denis Gorbunov}
\affiliation{Hochfeld-Magnetlabor Dresden (HLD-EMFL) and W\"urzburg-Dresden Cluster of Excellence ct.qmat, Helmholtz-Zentrum Dresden-Rossendorf, 01328 Dresden, Germany}

\author{Sergei Zherlitsyn}
\affiliation{Hochfeld-Magnetlabor Dresden (HLD-EMFL) and W\"urzburg-Dresden Cluster of Excellence ct.qmat, Helmholtz-Zentrum Dresden-Rossendorf, 01328 Dresden, Germany}

\author{Stanislaw Galeski}
\affiliation{Max Planck Institute for Chemical Physics of Solids, 01187 Dresden, Germany}

\author{Marcus Schmidt}
\affiliation{Max Planck Institute for Chemical Physics of Solids, 01187 Dresden, Germany}


\author{Jochen Wosnitza}
\affiliation{Institut f\"{u}r Festk\"orper- und Materialphysik, Technische Universit\"at Dresden, 01062 Dresden, Germany}
\affiliation{Hochfeld-Magnetlabor Dresden (HLD-EMFL) and W\"urzburg-Dresden Cluster of Excellence ct.qmat, Helmholtz-Zentrum Dresden-Rossendorf, 01328 Dresden, Germany}

\author{Johannes Gooth}
\email{johannes.gooth@cpfs.mpg.de}
\affiliation{Max Planck Institute for Chemical Physics of Solids, 01187 Dresden, Germany}
\affiliation{Institut f\"{u}r Festk\"orper- und Materialphysik, Technische Universit\"at Dresden, 01062 Dresden, Germany}

\date{\today}

\begin{abstract}
	In this study, we report on the observation of de Haas-van Alphen-type quantum oscillations (QO) in the ultrasound velocity of NbP as well as `giant QO' in the ultrasound attenuation in pulsed magnetic fields.
	The difference of the QO amplitude for different acoustic modes reveals a strong anisotropy of the effective deformation potential, which we estimate to be as high as $9\unit{eV}$ for certain parts of the Fermi surface.
	Furthermore, the natural filtering of QO frequencies and the tracing of the individual Landau levels to the quantum limit allows for a more detailed investigation of the Fermi surface of NbP as was previously achieved by means of analyzing QO observed in magnetization or electrical resistivity.
\end{abstract}
\maketitle
\section{Introduction}
Probing the propagation of ultrasound in the quantum regime of electrons yields detailed information on the nature of electron-phonon interactions.
%
%
The ultrasound velocity in such regime exhibits quantum oscillations (QO), which can be understood both from a thermodynamic argument \cite{Shoenberg,Testardi1971} as well as from a self-consistent treatment of ultrasound propagation as a stream of acoustic phonons interacting with an electron gas that is quantized into Landau levels (LL) \cite{Shapira1968,AcousticDHvA,Zhang2020,Mertschig1966}.
Both approaches yield the same result, namely, the amplitude of the QO being dependent on the (effective) deformation potential $\Xi_i^k=dE_k/d\varepsilon_i$, which is a measure for the change of energy $E_k$ of an electronic band $k$ at a given strain $\varepsilon_i$.
The connection to the microscopic picture can be understood intuitively by recalling that the probability for an electron in the $k$-th band of being scattered by a phonon-mode corresponding to $\varepsilon_i$ is proportional to $(\Xi_i^k)^2$ \cite{Bardeen1950,AcousticDHvA,Zhang2020,Shapira1968,Mertschig1966,Gurevich1961,Luethi2006}.
Employing measurements of magnetoacoustic QO, the deformation potential and its anisotropy have been experimentally determined for many metals and semimetals (see for example Refs.~\onlinecite{Shapira1968,Walther1968,Thompson1971,Matsui1995,Martins1978,Luethi2006,Noessler2017}).  
Recently, the semimetallic transition-metal monopnictide NbP is of great interest, mainly due to its symmetry-protected crossings of conduction and valence bands which potentially host Weyl fermions \cite{Weng2015,Xu2015,Huang2015}.
It exhibits a very small and highly anisotropic Fermi surface, consisting of intercalated spin-split pairs of electron and hole pockets due to spin-orbit coupling \cite{Klotz2016}.
The small Fermi surface gives rise to pronounced QO of relatively low frequencies, which have so far been observed in magnetization \cite{Stockert2017,Klotz2016,Sergelius2016}, electrical resistivity \cite{Shekhar2015,DosReis2016,Schindler2020,Niemann2017}, Hall resistivity \cite{Shekhar2015,Schindler2020}, thermal conductivity \cite{Stockert2017}, thermopower \cite{Stockert2017}, and heat capacity \cite{Stockert2017}.
%
%
The superposition of QO originating from different extremal Fermi-surface orbits yield a rich Fourier spectrum, especially when $\bm{H}$ is aligned along the $c$ axis of the tetragonal lattice and the extremal orbits are the smallest.
The peaks in the Fourier spectra could be assigned to orbits via comparison of experimental data to \emph{ab initio} density functional theory (DFT) calculations \cite{Klotz2016}, however, ambiguities due to the limited resolution and the broadness of the Fourier peaks remained.
In a recent study by some of the authors \cite{Schindler2020}, the evolution of the Fermi surface upon direct application of uniaxial stress along the $a$ axis has been probed by means of Shubnikov-de Haas (SdH) oscillations in the electrical resistivity.
These experiments revealed a strong strain dependence of the SdH oscillations, which, besides the additional information regarding the orbit assignments, also render NbP a promising platform for studying magnetoacoustic QO.
Furthermore, the strong anisotropy of the Fermi surface is suggestive of a highly anisotropic electron-phonon interaction as well, which can be most conveniently investigated via ultrasonic measurements.
In this paper, we report on the measurements of QO in the ultrasound velocity and attenuation in a NbP single crystal in pulsed magnetic fields $\bm{H}\parallel c$ (or [001]).
We have investigated the acoustic modes ($u\parallel q\parallel [100]$), ($u\parallel q\parallel [001]$), ($u\parallel [001], q\parallel [100]$), ($u\parallel [010], q\parallel [100]$), and ($u\parallel [1\bar{1}0], q\parallel [110]$) corresponding to the elastic moduli $C_{11}$, $C_{33}$, $C_{44}$, $C_{66}$, and $(C_{11}-C_{12})/2$ (using Voigt notation).
Here, $u$ is the displacement vector and $q$ is the direction of propagation of the acoustic wave.
Significant differences of the individual QO amplitudes between the modes were revealed.
A large signal-to-noise ratio, the usage of pulsed magnetic fields beyond the quantum limit, the high quality of our sample resulting in peak-shaped QO (presence of higher harmonics of the Fourier series), and the natural filtering of certain QO frequencies due to the anisotropic electron-phonon interaction allowed for a detailed analysis of the QO frequencies and amplitude ratios.
Thereby, the anisotropy of $\Xi_i^k$ and partially also the cyclotron masses, cyclotron mobilities, and phase factors for several extremal Fermi-surface orbits were determined.
The QO frequency spectrum could be analyzed via direct assignments of the LL peaks rather than Fourier analysis as in previous studies, which allowed for the assignment of formerly elusive orbits.
In addition, the extremal nature (maximum or minimum) of the individual orbits could be deduced from the asymmetric shape of the LL peaks.
\section{Methods}
NbP has a tetragonal crystal lattice (space group $I4_1md$, no. 109) with the lattice parameters  $a=b=3.3324(2)\,\si{\angstrom}$ and $c=11.13705(7)\,\si{\angstrom}$ \cite{Greenblatt1996}.
A single-crystalline sample of NbP was grown using chemical vapor transport reactions; the sample has also been used in our previous work \cite{Schindler2020} for the determination of the elastic moduli.
For acoustic modes propagating along one of the main axes, the sample was cut accordingly to a cuboid-shape of dimension $1.92\times 1.80\times 0.88\unit{mm^3}$.
For the $(C_{11}-C_{12})/2$ mode, two cuts parallel to the (110) plane were subsequently added.
The crystal planes were carefully polished and two lithium-niobate ($\mathrm{LiNbO_3}$) transducers (Z-cut for longitudinal waves and X-cut for transverse waves) were glued to opposite parallel surfaces for excitation and detection of acoustic waves.
The relative ultrasound-velocity changes $\Delta v/v$ and attenuation changes $\Delta \alpha$ were measured using an ultrasound pulse-echo phase-sensitive detection technique \cite{Wolf2001,Luethi2006} in pulsed magnetic fields up to $38\unit{T}$ (test pulses up to $56\unit{T}$) at temperatures ranging from 1.35 to $30\unit{K}$.
Excitation frequencies were varied from 27 to $100\unit{MHz}$ with pulse durations ranging from $50-200\,\mathrm{ns}$.
Strain-induction coupling, i.e., the Alpher-Rubin effect \cite{Testardi1971}, may be safely neglected at the used frequencies as the large magnetoresistance in NbP even at moderate magnetic fields ($\mu_0 H>1\unit{T}$) prevents from a strong skin effect. 
\section{Results}
\begin{figure}[ht!]
	\centering
	\includegraphics[width=8.6cm]{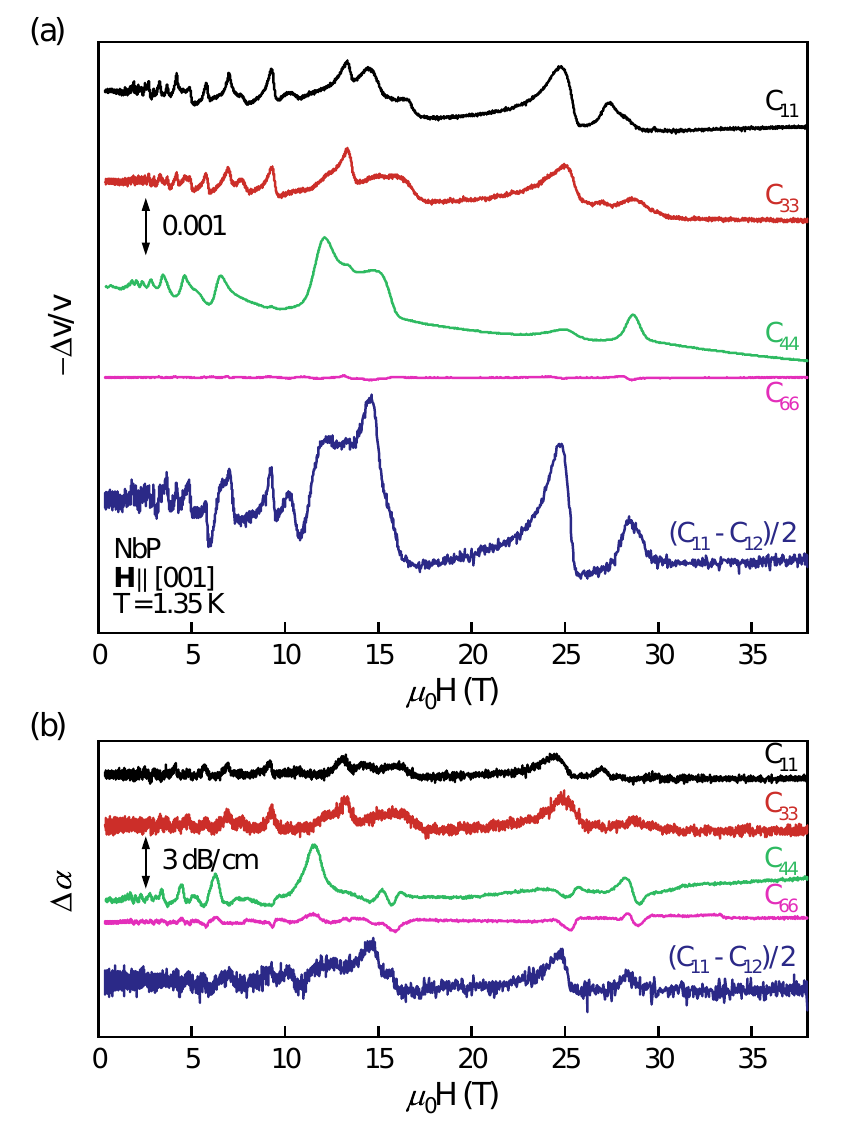}
	\caption{Magnetoacoustic quantum oscillations in NbP for pulsed magnetic fields $\bm{H}\parallel c$ at $T=1.35\unit{K}$ for different acoustic modes. (a) Change in the relative ultrasound velocity $-\Delta v/v$ versus magnetic field. (b) Change in ultrasound attenuation $\Delta\alpha$ versus magnetic field. The curves are shifted with respect to each other for better visibility.}
	\label{fig:field}
\end{figure}
The change of sound velocity $\Delta v/v$ and the change of sound attenuation $\Delta\alpha$ vs magnetic field at $T=1.35\unit{K}$ are shown for different acoustic modes in Fig.~\ref{fig:field}.
Here, $\Delta v/v$ refers to the change compared to the sound velocity at zero magnetic field $v=\sqrt{C_\mathrm{eff}/\rho}$, where $C_\mathrm{eff}$ is the effective elastic constant governing the respective mode \cite{Brugger1965} and $\rho$ is the mass density ($\rho=6.52\,\mathrm{g\,cm^{-3}}$ for NbP \cite{Greenblatt1996}).
$\Delta v/v$ shows pronounced QO with high harmonic content, whereas dominant frequencies and size of the oscillation amplitudes strongly vary between the modes.
Strikingly, the QO amplitude in the $C_{66}$ mode is smaller by a factor of $\approx 20$ compared to the other modes, where for the last few LL changes in $v$ by more than one part in a thousand are observed.
$\Delta\alpha$ exhibits QO with a characteristic spike-like shape, also varying in terms of amplitude and dominant frequencies depending on the mode.
We recall that the physical mechanism responsible for the QO in ultrasound attenuation, which are commonly termed as `giant QO' \cite{Gurevich1961,Shoenberg}, is not related to the Landau tubes passing through the extremal parts of the Fermi surface as in the de Haas-van Alphen (dHvA)-type oscillations.
Instead, spikes in $\Delta\alpha$ occur when the Landau tubes pass through the Fermi-surface section, where the component of the Fermi velocity parallel to $q$ is equal to the phase velocity of sound \cite{Gurevich1961,Shoenberg,Shapira1968,Mertschig1970}.
This resonance condition is the reason for the spike-like shape, as it is only fulfilled for particular values of the wavevector in contrast to the contribution of many wavevectors in the dHvA-type oscillations.
Notably, the resonant Fermi-surface orbits can differ substantially from the extremal orbits, especially when $q\perp\bm{H}$. 
Hence, the position of the observed spikes in $\Delta\alpha$ do not necessarily coincide with the LL peaks in $\Delta v/v$.
Above $30\unit{T}$, all electrons and holes are confined to their lowest LL; and $v(H)$ and $\alpha(H)$ exhibit a steady slope in the investigated field (measured up to $56\unit{T}$ for $C_{44}$) and temperature range, showing no signatures for correlation-driven charge instabilities.
Such correlation-driven phase transitions, e.g., a charge density wave, would manifest in a slope change of $\Delta v/v$ and a peak in $\Delta\alpha$ \cite{LeBoeuf2017}, and have been predicted to occur in the extreme quantum limit of Weyl semimetals \cite{Trescher2017,Laubach2016}.
Notably, there have been observations of indicative features in the extreme quantum limit in the electrical resistivity and in the sound velocity and attenuation in the related compound TaAs \cite{Ramshaw2018,Zhang2016}.
However, in case of pristine NbP the interaction strength presumably is too feeble as to allow for experimental access to these energy scales within our achievable field and temperature range.
\subsection{Quantum oscillations in the velocity of sound}
\subsubsection{Frequency analysis and orbit assignment}
\begin{figure*}[ht!]
	\centering
	\includegraphics[width=17.2cm]{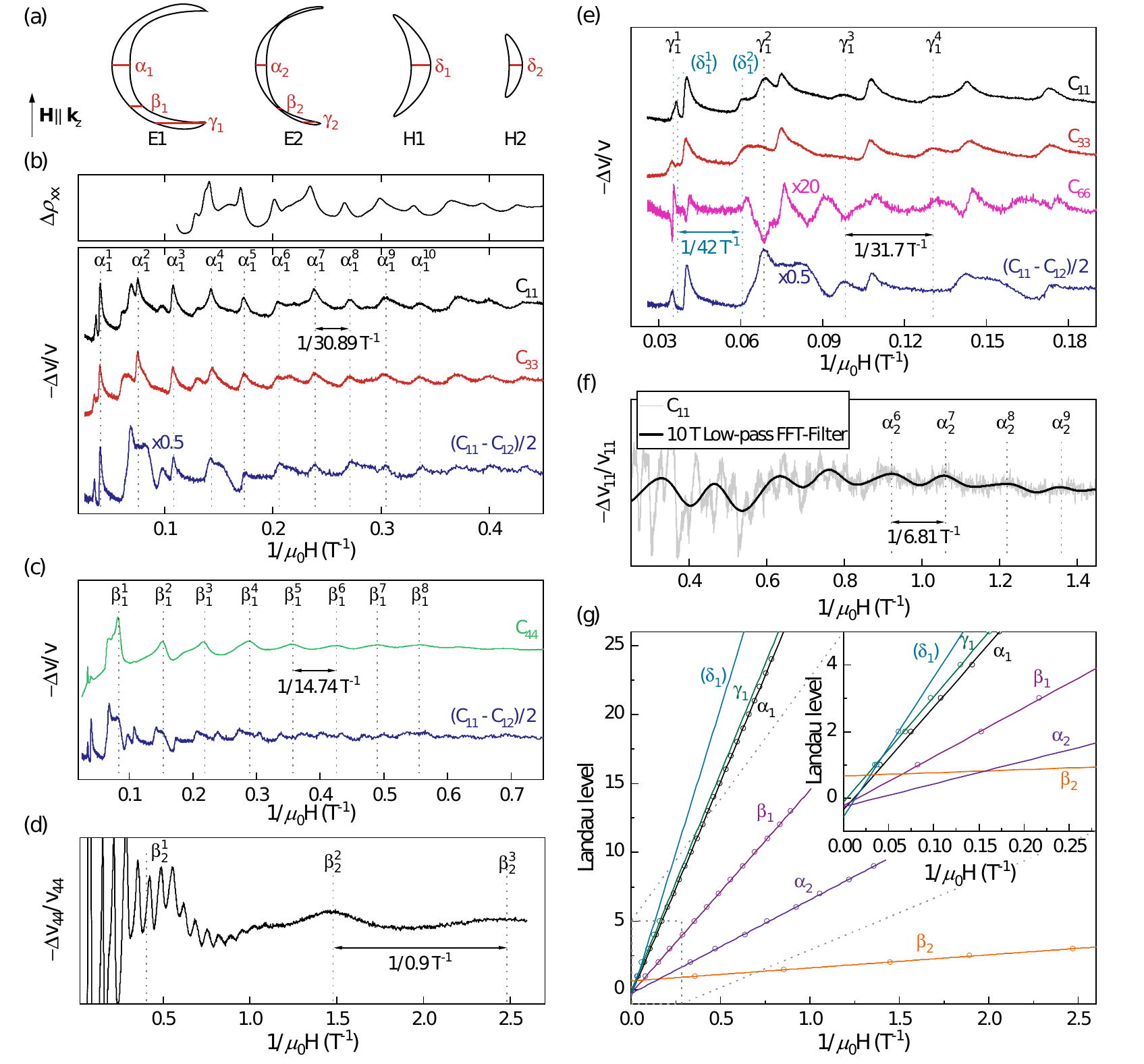}
	\caption{Frequency analysis of the quantum oscillations in ultrasound velocity for different modes at $T=1.35\unit{K}$. (a) Projections of the electron pockets, E1 and E2, and the hole pockets, H1 and H2, of NbP parallel to the $k_x$-$k_z$ plane (or similarly to the $k_y$-$k_z$ plane due to the fourfold rotational symmetry). Extremal orbits for $\bm{H}\parallel c$ are shown in red. For an illustration of the full Fermi surface in the first Brillouin zone see for instance Refs.~\onlinecite{Schindler2020,Lee2015}. (b) Top: Shubnikov-de Haas oscillations subtracted from the magnetoelectrical resistivity $\rho_{xx}$ at $T=2\unit{K}$ for comparison. Bottom: Landau-level peaks assigned to the maximum orbit $\alpha_1$. (c) Landau-level peaks assigned to the minimum orbit $\beta_1$. (d) Low-frequency oscillation visible in the $C_{44}$ mode assigned to the minimum orbit $\beta_2$. (e) Assignment of the remaining peaks in the high-field range to the second maximum orbit of E1, $\gamma_1$, and possibly the maximum orbit $\delta_1$. (f) Oscillation assigned to the maximum orbit $\alpha_2$ visible in the $C_{11}$ mode, emphasized by applying a low-pass Fourier filter. (g) Assigned Landau levels plotted versus inverse magnetic field. Solid lines represent linear fits. The inset enlarges the high-field range.}
	\label{fig:frequencies}
\end{figure*}
\begin{table*}[ht!]
	\caption{Experimental results extracted from the analysis of quantum oscillations in the ultrasound velocity for $\bm{H}\parallel c$. The calculated orbits are denoted as in Ref.~\onlinecite{Klotz2016} and experimentally extracted frequencies $F$ are assigned as in Refs.~\onlinecite{Klotz2016,Schindler2020}, considering also the asymmetry of the Landau level peaks due to the extremal nature of the orbit (maximum or minimum). The cyclotron mass $m_\mathrm{c}$, cyclotron mobility $\mu_\mathrm{c}$, Dingle temperature $T_\mathrm{D}$, phase factor $\varphi$ and the effective deformation potential $\Xi_i$ with respect to $\Xi_1$ are given if possible (absolute values only). $\Xi_\mathrm{s}$ denotes the deformation potential corresponding to the $(C_{11}-C_{12})/2$ mode.}
	\label{tab:Data}
	\begin{ruledtabular}
		\begin{tabular}{c|l|c|c|c|c|c|c|c|c|c|c|c}
			Orbit & Extr. & $F_\mathrm{theo} (\mathrm{T}) \footnote{The calculated frequencies were obtained from density functional theory in our previous study \cite{Schindler2020}.}$ & $F_\mathrm{exp}(\mathrm{T})$ & $m_\mathrm{c}(m_0)$ & $\mu_\mathrm{c} (\mathrm{10^{3} cm^2 V^{-1} s^{-1}})$ & $T_\mathrm{D}(\mathrm{K})$ & $\varphi$ & $\Xi_1(\mathrm{eV}$) \footnote{$\Xi_1$ has been estimated with Eq.~(\ref{eq:deformation}) using the averaged $\partial F/\partial \varepsilon_1$ values from Ref.~\onlinecite{Schindler2020}.} & $\Xi_3/\Xi_1$ & $\Xi_4/\Xi_1$ & $\Xi_6/\Xi_1$ & $\Xi_\mathrm{s}/\Xi_1$ \\ \hline
			\multicolumn{13}{c}{Electron pocket E1} \\ \hline
			$\alpha_1$ & Max & 32.8 & 30.89(5) & 0.06(1) & 25(5)& 1.4(6) & 0.27(1) & 2.1(5) & 0.9(1) & 0.7(1) & 0.24(4) & 2.0(2) \\
			$\beta_1$ & Min & 11.3 & 14.74(4) & 0.12(2) & 9(1) & 2.0(5) & 0.23(1) & 1.4(3) & 1.2(1) & 6.3(5) & 0.8(1) & 5.1(4) \\
			$\gamma_1$ & Max & 31.1 & 31.7(5) & - & - & -  & 0.20(2) & -  & 1.6(1) & 3.0(2) & 0.6(1) & 3.2(3) \\ \hline
			\multicolumn{13}{c}{Electron pocket E2} \\ \hline
			$\alpha_2$ & Max & 7.92 & 6.81(7) & - & - & - & 0.5(1) & - & - & - & - & - \\
			$\beta_2$ & Min & $\approx 1$ & 0.9(1) & 0.022(4) & 70(20) & 1.4(5) & 0.49(2) & - & - & - & - & - \\
			$\gamma_2 $ & Max & 4.7 & - & - & - & - & - & - & - & - & - & - \\ \hline
			\multicolumn{13}{c}{Hole pocket H1} \\ \hline
			$\delta_1 $ & Max & 41.4 & 42(1) & - & - & - & - & - & 0.8(1) & $\approx 0$ & $\approx 0$ & $\approx 0$ \\ \hline
			\multicolumn{13}{c}{Hole pocket H2} \\ \hline
			$\delta_2 $ & Max & 22.1 & - & - & - & - & - & - & - & - & - & - \\
		\end{tabular}
	\end{ruledtabular}
\end{table*}
To analyze the QO in the ultrasound velocity, $-\Delta v/v$ is plotted against $1/H$ (Fig.~\ref{fig:frequencies}).
The ultrasound velocity, just as any thermodynamic property of a material, exhibits singularities upon increasing magnetic field whenever a cyclotron orbit corresponding to a LL is exactly equal to an extremal orbit of the Fermi-surface sheet perpendicular to the applied $\bm{H}$.
According to the Onsager relation \cite{Shoenberg}, these singularities are periodic in $1/H$ with the frequency $F=(\hbar/2\pi e)A_\mathrm{ext}$, where $A_\mathrm{ext}$ is the area enclosed by the corresponding extremal orbit, $\hbar$ is the reduced Planck constant and $e$ the electron charge.
Plotting LL number vs $1/H$, $F$ can then be extracted using a linear fit [see Fig.~\ref{fig:frequencies}(g)].
For a maximum orbit, $-\Delta v/v$ will increase with $(1/H)^{-1/2}$ approaching a LL singularity from a lower field and then decrease steeply, once the area of the corresponding cyclotron orbit exceeds that of the maximum orbit \cite{Mertschig1970}.
Accordingly, for a minimum orbit these slopes are reversed and the steep rise appears on the low-field side of the LL peak.
If smearing due to finite temperature and electron scattering is sufficiently suppressed, the QO retain a high harmonic content and approach a sawtooth-like shape.
The asymmetry of the individual LL peaks then allows for identifying whether the corresponding peak is arising from a maximum or minimum orbit of the Fermi surface.
Clearly, the dominant frequency of $30.89\unit{T}$ in $C_{11}$ and $C_{33}$ (also very well distinguishable in the $(C_{11}-C_{12})/2$ mode) stems from a maximum orbit [most apparent for the last three LL, see Fig.~\ref{fig:frequencies}(b)].
It is also the most pronounced frequency in the SdH oscillations in magnetoresistance [Fig.~\ref{fig:frequencies}(b) top], whose shape resembles that of the $C_{11}$ mode.
As assigned in Ref.~\onlinecite{Klotz2016} based on DFT calculations and further indicated by comparing experimental and calculated strain dependences \cite{Schindler2020}, this frequency is likely stemming from the $\alpha_1$ rather than the $\gamma_1$ orbit [hereafter, we use the same labeling for the extremal orbits of NbP as in these Refs., see Fig.~\ref{fig:frequencies}(a)].
The $\alpha_1$ oscillation is much less pronounced in $C_{44}$ [see Fig.~\ref{fig:frequencies}(c)], allowing for a clear identification of the $14.74\unit{T}$ oscillation as a minimum orbit, assigned to $\beta_1$.
After having identified the LL peaks for $\alpha_1$ and $\beta_1$, the remaining peaks in the high-field range might be assigned to the $\gamma_1$ orbit and possibly also the $\delta_1$ orbit [see Fig.~\ref{fig:frequencies}(e)].
The assignment to $\delta_1$ is thereby rather speculative; the second peak at approx. $0.06\unit{T^{-1}}$ might also stem from the last LL of $\delta_2$.
At low fields, a $0.9\unit{T}$ oscillation with minimum-orbit characteristics is visible in $C_{44}$, assigned to $\beta_2$ [Fig.~\ref{fig:frequencies}(d)].
Furthermore, by applying a low-pass Fourier filter to $C_{11}$ an oscillation of $6.81\unit{T}$ is singled out, which was also identified in the Fourier spectra from previous QO studies \cite{Klotz2016,DosReis2016,Schindler2020} and assigned to the $\alpha_2$ orbit [Fig.~\ref{fig:frequencies}(f)].
The extracted frequencies are summarized in Table~\ref{tab:Data}.
We note that we did not observe additional QO patterns predicted to occur in Weyl semimetals when the Fermi level is near the Weyl points \cite{Zhang2020}. 

\subsubsection{Lifshitz-Kosevich fit}
\begin{figure*}[]
	\centering
	\includegraphics[width=17.2cm]{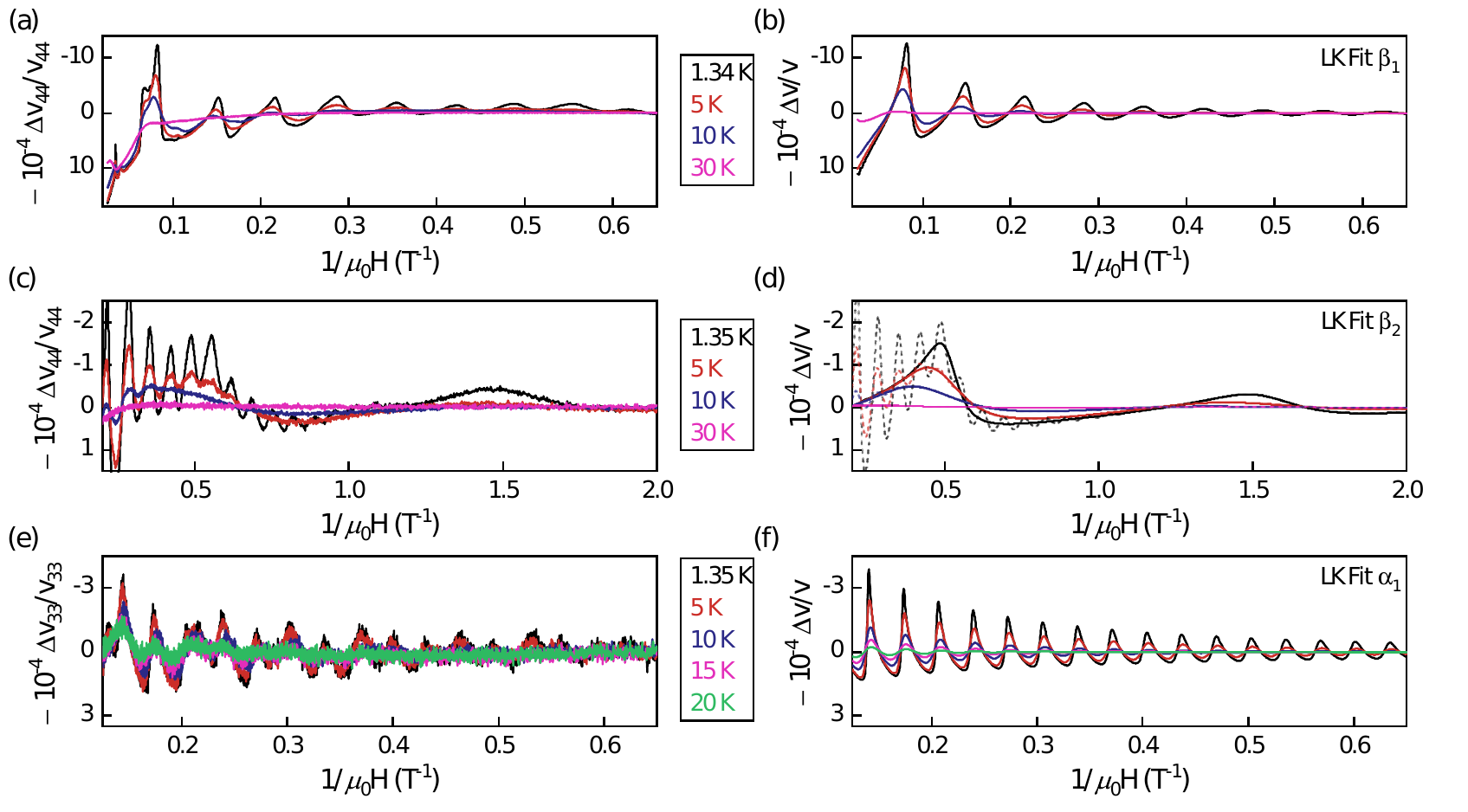}
	\caption{Temperature evolution of the quantum oscillations in the ultrasound velocity for the $C_{44}$ [(a) and (c)] and $C_{33}$ mode (e) and Lifshitz-Kosevich fit for the frequencies $\beta_1$ (b), $\beta_2$ (d), and $\alpha_1$ (f) dominant in $C_{44}$ and $C_{33}$, respectively.}
	\label{fig:temperature}
\end{figure*}
The actual shape of the QO in $\Delta v/v$ can be described by a Fourier series taking finite-temperature smearing of the Fermi-Dirac distribution and LL broadening due to electron scattering into account.
After Lifshitz and Kosevich \cite{Shoenberg}, the oscillatory part of $\Delta v/v$ for a single QO frequency without spin degeneracy holds
\begin{align}
	\frac{\tilde{v}_{ij}}{v_{ij}} = - \frac{1}{2} \left(\frac{\partial F}{\partial \varepsilon_i}\right) \left(\frac{\partial F}{\partial \varepsilon_j}\right)\frac{e^2 V}{m_\mathrm{c}C_{ij}}\left(\frac{2eH}{\hbar\pi^3A_\mathrm{ext}"}\right)^{\frac{1}{2}} \nonumber \\
	\times \sum_{p=1}^{\infty}p^{-\frac{1}{2}}R_TR_\mathrm{D}\cos\left[2\pi p\left(\frac{F}{H}-\varphi\right)\pm\frac{\pi}{4}\right], \label{eqn:1}
\end{align}
where $m_\mathrm{c}$ denotes the effective cyclotron mass, $V$ the real space volume, $A_\mathrm{ext}"$ is the curvature of the Fermi surface at the extremal orbit and $\varphi$ is the phase factor.
The $\pm\pi/4$ phase shift accounts for whether the orbit is maximum ($-$) or minimum ($+$).
Damping of the QO due to thermal smearing of the Fermi distribution is accounted for by the factor \cite{Shoenberg}
\begin{align}
	R_\mathrm{T} = \frac{\lambda(T)}{\sinh\left[\lambda(T)\right]},~\mathrm{with}~\lambda(T)=p\frac{2\pi^2m_\mathrm{c}k_\mathrm{B}T}{e\hbar H}.
\end{align}
Damping due to electron scattering is taken into account by the Dingle damping factor \cite{Shoenberg}
\begin{align}
	R_\mathrm{D}=\exp\left[-\lambda(T_\mathrm{D})\right]=\exp\left[-p\frac{\pi}{\mu_\mathrm{c}H}\right],
\end{align}
where $T_\mathrm{D}$ is the Dingle temperature and $\mu_\mathrm{c}$ is the mobility of an electron exerting a cyclotron motion in an applied magnetic field (not to be confused with the zero-field transport mobility, which, depending on the current direction, can significantly differ from $\mu_\mathrm{c}$ in case of a large band anisotropy \cite{pippard1989magnetoresistance}).
The $\beta_1$, $\beta_2$, and $\alpha_1$ oscillations were clearly distinguishable in $C_{44}$ and $C_{33}$, respectively, and could be approximated using the first 20 harmonics of Eq.~(\ref{eqn:1}).
From fits to the QO for different temperatures (Fig.~\ref{fig:temperature}), the damping factors $R_\mathrm{D}$ and $R_T$ could be extracted, allowing for the determination of $m_\mathrm{c}$, $\varphi$, $\mu_\mathrm{c}$, and $T_\mathrm{D}$ (summarized in Table~\ref{tab:Data}).
The fitting procedure was performed globally for all temperatures with the shared parameters $F$ (fixed), $m_\mathrm{c}$, $\varphi$, and $\mu_\mathrm{c}$, and an independent amplitude prefactor.
We note that the direct fitting of the naturally filtered QO yields a greater reliability for the $m_\mathrm{c}$ values compared to the analysis of Fourier spectra, as there the field-dependent amplitude damping usually leads to a systematic underestimation of $m_\mathrm{c}$ \cite{Schindler2020,Audouard2018}.
Our fits yielded an effective cyclotron mass of $0.06(1)m_0$ for $\alpha_1$ and $0.12(2)m_0$ for $\beta_1$, which is larger than the values extracted from Fourier analysis of dHvA oscillations \cite{Klotz2016} [$0.047(9)m_0$ and $0.057(7)m_0$].
The extracted $m_\mathrm{c}$ are also in better agreement with the calculated values from Ref.~\onlinecite{Klotz2016} ($0.10m_0$ and $0.12m_0$) compared to previous methods, albeit this does not necessarily imply improved accuracy.
\subsubsection{Discussion of the phase factor}
The phase factors extracted from fitting Eq.~(\ref{eqn:1}) to the $\Delta v/v$ data are around $0.5$ for the extremal orbits $\alpha_2$ and $\beta_2$ on the electron pocket E2, and vary from $0.27$ to $0.20$ for the orbits $\alpha_1$, $\beta_1$ and $\gamma_1$ on E1.
According to recent theory works by Alexandradinata et al. \cite{Alexandradinata2018,Alexandradinata}, the phase factor generally consists of three contributions 
\begin{align}
	\varphi = \varphi_\mathrm{M}- \varphi_\mathrm{B} - \varphi_\mathrm{d},
\end{align}
where $\varphi_\mathrm{M}$ is the Maslov correction ($\varphi_\mathrm{M}=1/2$ for orbits that are compressible to a circle, which is the case for all orbits in NbP), $\varphi_\mathrm{B}$ is the geometric phase, i.e., Berry phase \cite{BerryPhase1999}, that an electron acquires upon encircling the orbit in reciprocal space, and $\varphi_\mathrm{d}$ is the dynamic phase factor which accounts for the generalized Zeeman interaction of the intrinsic and orbital magnetic moment.
The main interest in analyzing the phase contributions lies in the extraction of $\varphi_\mathrm{B}$, as it potentially allows to identify topologically non-trivial bands, such as Weyl or Dirac bands \cite{BerryPhase1999}.
Indeed, under certain symmetry constraints (for details, see Refs.~\onlinecite{Alexandradinata2018,Alexandradinata}) $\varphi_\mathrm{d}$ vanishes or can only take quantized values $\pm1/2$, which then allows to draw conclusions about $\varphi_\mathrm{B}$.
As all orbits in NbP for $\bm{H}\parallel c$ can be mapped onto themselves in $k$ space upon applying a mirror operation (mirror planes $k_x=0$ or $k_y=0$, see Ref.~\onlinecite{Lee2015}), they belong to the classification ($\mathrm{II}$-$\mathrm{A}$, $u=1$, $s=0$) of Tab.~I in Ref.~\onlinecite{Alexandradinata2018}, and $\varphi_\mathrm{d}$ can be either $0$ or $1/2$ depending on details of the band structure.
Hence, at first glance it may seem that a deviation of $\varphi$ from $0$ or $1/2$ can be regarded as a signature of a non-zero $\varphi_\mathrm{B}$.
However, it was shown by Klotz \emph{et al.} \cite{Klotz2016} that the Fermi-surface pockets in NbP intersecting with the Weyl bands, E1 and H1, always encompass a pair of Weyl points and should thus exhibit a trivial phase shift of $\varphi_\mathrm{B}=1$ or 0.
Hence, the extracted phase factors of $\alpha_1$, $\beta_1$ and $\gamma_1$ are at odds with the possible values predicted by theory.
It is rather speculative why this is the case, the reason might be slight misalignment of the magnetic field, wrong orbit assignment or, more generally, inaccuracy of the DFT calculations, although the latter two are highly improbable given the otherwise good agreement.
The extracted $\varphi$ of E2 are not contradicting theory, but are also not particularly informative regarding the topological nature of the bands.
%
%
%
%
%
%
%
%
%
%
%

%
%
\subsubsection{Extraction of the deformation potentials}
\begin{figure}[]
	\centering
	\includegraphics[width=8.6cm]{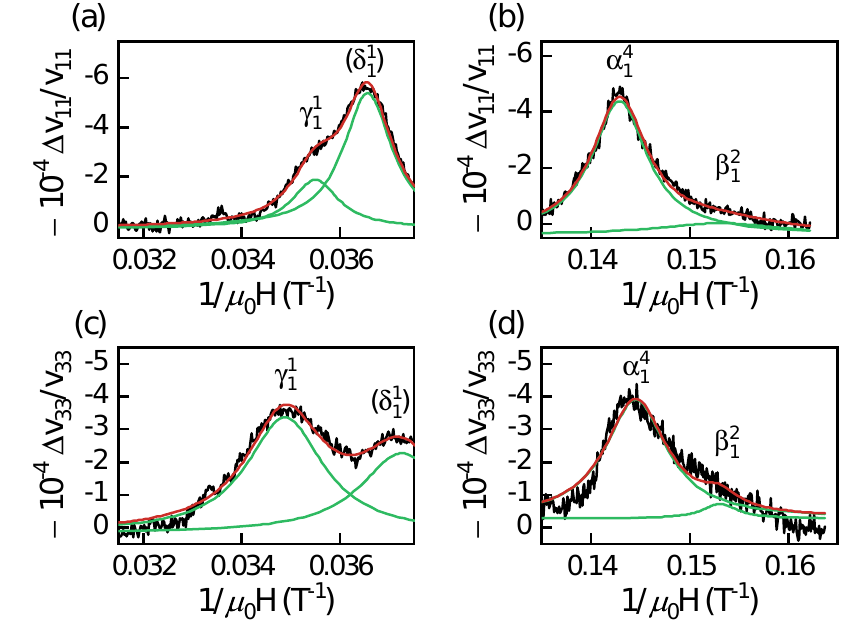}
	\caption{Extraction of the oscillation amplitude for superimposed peaks in the high-field range by fitting two Lorentzian functions (green) at fixed inverse-field values. (a) and (c) Extraction of the height for $\gamma_1^1$ and $\delta_1^1$ from the ultrasonic quantum oscillations in the $C_{11}$ (a) and $C_{33}$ mode (c). (b) and (d) Extraction of the height for $\beta_1^2$ for the $C_{11}$ (b) and $C_{33}$ mode (d).}
	\label{fig:amplitude}
\end{figure}

Comparing the amplitudes of the same orbit for different modes, the ratio of the $C_{ii}^{-1}(dF/d\varepsilon_i)^2$ values can be extracted.
With the known elastic constants from our previous study \cite{Schindler2020}, the ratio of the effective deformation potentials can then be calculated via \cite{Testardi1971}
\begin{align}
	\Xi_i=\frac{dE}{d\varepsilon_i}=\frac{dE}{dA_\mathrm{ext}}\frac{dA_\mathrm{ext}}{d\varepsilon_i}=\frac{\hbar e}{m_\mathrm{c}}\frac{\partial F}{\partial \varepsilon_i}.
	\label{eq:deformation}
\end{align}
The amplitude ratios for the individual orbits have been extracted by selecting well distinguishable LL peaks (near the quantum limit) and divide their top-to-bottom heights.
In case there was no separate LL peak, as for example for the $\beta_1$ orbit in $C_{11}$ and $C_{33}$, the height was estimated via fitting of two Lorentzian functions with fixed centers (Fig.~\ref{fig:amplitude}), whereas the center positions were extracted from comparison with other modes (see Fig.~\ref{fig:frequencies}).
The resulting deformation potentials w.r.t. $\Xi_1$ are summarized in Table~\ref{tab:Data}.
They are strongly anisotropic - measurable $\Xi$ values vary by up to a factor of $\approx8$ depending on the direction of strain - which reflects the anisotropy of the electronic bands in NbP (see DFT calculations in Refs.~\onlinecite{Lee2015,Sun2015}).
In contrast to the isotropic behavior in conventional metals, the electron-phonon scattering in NbP [and transferably other (Weyl) semimetals with anisotropic bands] is highly selective.
With the $\partial F/\partial \varepsilon_1$ values gathered from Ref.~\onlinecite{Schindler2020}, $\Xi_1$ can be estimated via Eq.~(\ref{eq:deformation}) to be $2.1\unit{eV}$ ($2.5\unit{eV}$) for $\alpha_1$ and $1.4\unit{eV}$ ($2.2\unit{eV}$) for $\beta_1$ taking experimentally (calculated) values, respectively.
For $\beta_1$, this results in an effective deformation potential of $9\unit{eV}$ ($14\unit{eV}$) for shear strain along $c$.
This potential is among the highest reported values \cite{Walther1968,Tekippe1972,Virgaftman2001} and illustrates how electrons in the narrow part of the electron pocket are extremely susceptible to interaction with phonon modes corresponding to such shear strain.
We note that upon applying strain along an axis perpendicular to the $c$ axis, the breaking of the rotational symmetry leads to a degeneracy lifting of the Fermi pockets and $\partial F/\partial \varepsilon_1$ actually splits into a positive and a negative branch \cite{Schindler2020}.
As in Eq.~(\ref{eqn:1}) the sign of $\partial F/\partial \varepsilon_1$ is canceled due to the square, we took the average of the absolute values in order to estimate $\Xi_1$.
\subsection{`Giant' quantum oscillations in ultrasound attenuation}
\begin{figure}[h!]
	\centering
	\includegraphics[width=8.6cm]{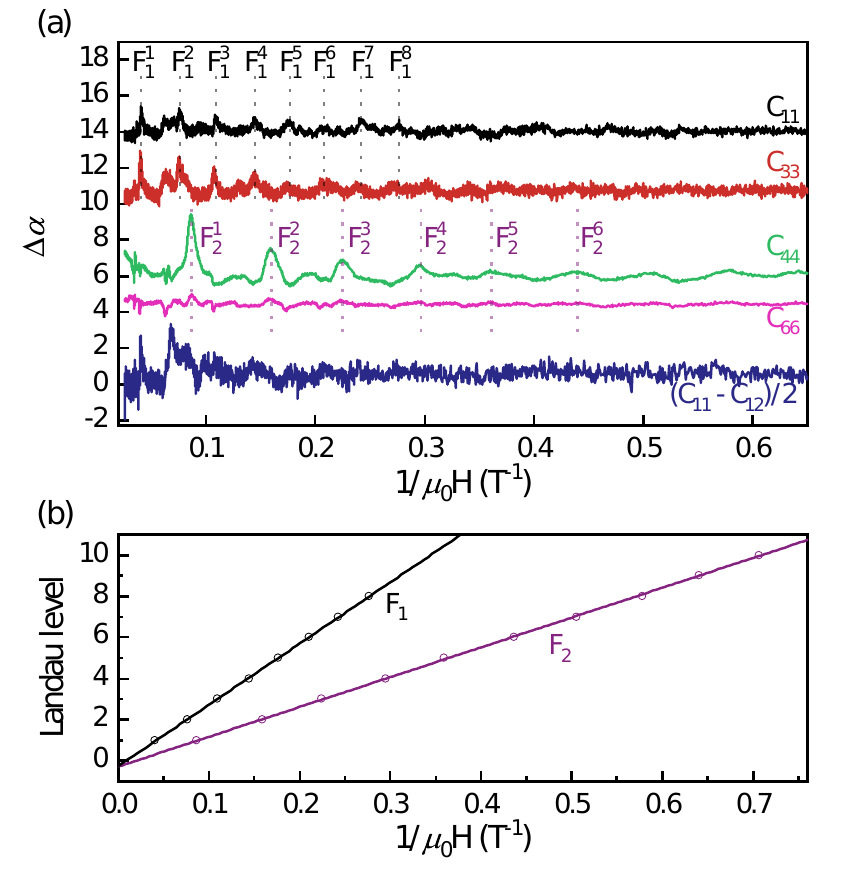}
	\caption{Frequency analysis of the `giant quantum oscillations' in ultrasound attenuation for different modes at $T=1.35\unit{K}$. (a) Landau level peaks assigned to the resonant orbits $F_1$ and $F_2$. (b) Assigned Landau levels plotted vs inverse magnetic field. Solid lines represent linear fits.}
	\label{fig:attenuation}
\end{figure}
The `giant QO' in $\Delta\alpha$ are less straightforward to analyze, as the position of the resonant orbits in reciprocal space is rather complicated to determine for each corresponding phonon mode.
If plotted against $1/H$ [Fig.~\ref{fig:attenuation}(a)], two periodic series of spikes are very well distinguishable, labeled as $F_1$ and $F_2$.
The Onsager relation is valid for the `giant QO' as well; linear fits to the spike positions vs LL number yield $F_1=29.8\unit{T}$ and $F_2=14.5\unit{T}$ [Fig.~\ref{fig:attenuation}(b)].
The areas enclosed by the resonant orbits are thus close to those of $\alpha_1$ and $\beta_1$.
A puzzling feature is the observation of the same frequencies in two modes with perpendicular $q$, e.g., $F_1$ in both $C_{11}$ and $C_{33}$.
This observation might be explained by the peculiar shape of the Fermi surface in NbP, where fourfold degenerate sickle-like pockets are located near the edges of the first Brillouin zone.
In this particular case, the resonant condition might be fulfilled for the same orbit for elastic waves propagating both along $a$ and $c$.
In contrast to the QO in $\Delta v/v$, the exact shape of the spikes in $\Delta\alpha$ is rather difficult to fit.
Each $\delta$ function corresponding to a spike must be convoluted with various distribution functions accounting for the effects of finite temperature and electron scattering \cite{Shoenberg}.
In our case, this did not seem viable as multiple frequencies superimpose each other and similar information on the electronic properties has already been extracted from the QO in $\Delta v/v$, where also the signal-to-noise ratio was more favorable.
Nevertheless, the slight asymmetry of the spikes can be attributed to an indirect effect of electron scattering, where the smearing of the LL relaxes the resonance condition \cite{Shoenberg}.
The spikes of $F_1$ and $F_2$ are broader towards the low-field side [see Fig.~\ref{fig:field}(b)], which is indicative of a convex curvature of the Fermi surface at the resonant orbit ($A"<0$).
\section{Summary}
In summary, we studied the QO in ultrasound velocity and attenuation in NbP in pulsed magnetic fields.
Thereby, fields with $\bm{H}\parallel c$ beyond the quantum limit were applied.
We compared the QO for several acoustic modes, revealing significant differences as to which orbits are dominant.
By extracting the amplitudes of the QO in the ultrasound velocity, the anisotropy of the deformation potentials has been determined for several extremal orbits.
Thereby, a large deformation potential of approximately $9\unit{eV}$ for the minimum orbit $\beta_1$ under shear strain along the $c$ axis has been revealed, suggesting that electrons in this part of the Fermi surface are very susceptible to interactions with the phonon modes corresponding to $C_{44}$.
Furthermore, the high harmonic content of the QO and the large field range allowed for a more reliable determination of the frequencies, effective cyclotron masses, and mobilities as was previously achieved by means of Fourier analysis.
On a side note, we did not find any signatures for correlated electron states in the quantum limit of (pristine) NbP.
%
%
%
%
%
%
%
%
%
%
%
%
%
%
%
\begin{acknowledgments}
	C.~S. would like to thank A. Alexandradinata for engaging in helpful discussions.
	C.~S. acknowledges financial support by the International Max Planck Research School for Chemistry and Physics of Quantum Materials (IMPRS-CPQM).
	%
	%
	The work was supported by Deutsche Forschungsgemeinschaft (DFG) through SFB 1143 and the W\"urzburg-Dresden Cluster of Excellence on Complexity and Topology in Quantum Matter---$ct.qmat$ (EXC 2147, project-id 390858490), and by Hochfeld-Magnetlabor Dresden (HLD) at HZDR, member of the European Magnetic Field Laboratory (EMFL).
	%
\end{acknowledgments}
%
%
%
%
\providecommand{\noopsort}[1]{}\providecommand{\singleletter}[1]{#1}%

\end{document}